\documentstyle[pre,aps,epsfig,twocolumn]{revtex}

\def\break#1{\pagebreak \vspace*{#1}}

\begin{document}

\addtolength {\oddsidemargin} {-1cm}
\addtolength {\topmargin} {1cm}
\setlength{\parindent}{0.4cm}

\title{Evidence of universality for the May-Wigner stability theorem
for random networks with local dynamics}

\author{Sitabhra Sinha{\footnote{Electronic address: sitabhra@imsc.res.in}} and
Sudeshna Sinha{\footnote{Electronic address: sudeshna@imsc.res.in}} }

\address{The Institute of Mathematical Sciences, C. I. T. Campus,
Taramani, Chennai - 600 113, India.}

\maketitle

\vspace{-1cm}
\widetext
\begin{abstract}
  We consider a random network of nonlinear maps exhibiting a wide
  range of local dynamics, with the links having normally distributed
  interaction strengths. The stability of such a system is examined in
  terms of the asymptotic fraction of nodes that persist in a non-zero
  state. Scaling results show that the probability of survival in the
  steady state agrees remarkably well with the May-Wigner stability
  criterion derived from linear stability arguments. This suggests
  universality of the complexity-stability relation for random
  networks with respect to arbitrary global dynamics of the system.
\end{abstract}

\vspace{-0.4in}
\narrowtext 
The relation between the structure of a network and its
dynamical properties has been a problem of long-standing importance in
many fields, especially in theoretical ecology \cite{Dro02}. A major
advance in this area was the suggestion by May that the stability of a
network can be inferred from an analysis of the interactions between
the network elements \cite{May73}. Confining attention only to the
local stability of an arbitrary equilibrium of the dynamics, one can
ignore explicit dynamics and look at only the leading eigenvalues of
the linear stability matrix.  Assuming that the network interactions
are random, rigorous results on the eigenvalue spectra of random
matrices can be applied \cite{Meh91}.  If the stability matrix is
comprised of elements from a normal distribution with zero mean and
variance $\sigma^2$, then the network is almost certainly stable if $N
C \sigma^2 < 1$, and unstable otherwise.  $N$ is the number of nodes
in the network and $C$ is the network connectivity, i.e., the
probability that any two given elements of the network are coupled to
each other, as reflected in the sparsity of the matrix
\cite{May73,May72}. This result is often referred to as the May-Wigner
stability theorem \cite{Has82}.

May's suggestion that increasing network complexity leads to
decrease in stability was supported by earlier numerical
simulations \cite{Gar70}, but it ran counter to the empirically established
conventional wisdom that biodiversity promotes ecosystem stability.
The original result has been criticized on the
ground that it is obtained by linearizing about an assumed
equilibrium, and so is inapplicable when either the perturbations from
the equilibrium are large, or, the dynamics does not settle down to a
fixed point attractor (e.g., they 
might undergo periodic oscillations as in a Lotka-Volterra type system).
The ensuing stability vs diversity debate in ecology has resulted in a
large body of literature attempting to resolve this issue one way or
another \cite{McC00}. Although much of the controversy may have been
due to the methods that different groups used to measure complexity
and stability \cite{Pim84}, and the two apparently opposing
conclusions have been resolved in the specific context of a community
assembly model \cite{Wil02}, the general question of whether network
complexity is conducive to the long-term persistence of the nodes
remains unresolved.  
In addition to ecological networks, phenomena where the survival
\break{0.95in}
of nodes in a network maybe of relevance are power grid breakdown, 
financial market crashes, etc., in short, any
system that is susceptible to sudden collapse.
Further, since the present problem is related to the persistence of a 
trajectory in a high-dimensional space with absorbing boundaries,
it is also of considerable relevance to the general question of persistence 
in non-equilibrium systems that has seen a huge spurt of
interest recently \cite{Maj99}. 

In this paper, we report results on the role that network complexity
plays on global stability (in contrast to local stability) of a
network, by looking at the persistence of individual nodes in a
network of randomly coupled nonlinear maps undergoing a wide range of
local dynamics.  We observe that the results of the May-Wigner theorem
seem to be valid universally, namely, increasing the number of interactions
per node or increasing interaction strength will give rise to increased
likelihood of extinction. This evidence of universality (in the sense of 
being independent of the local dynamics at the nodes) has bearing on
network problems in general \cite{Rag95,Wat02}, as it addresses an
issue which arises in many different contexts, namely: {\em what is
the significance of local dynamics on network stability, especially
in situations where the dynamics can be widely varying}.

Previous work on including explicit dynamics in network models mostly
involved generalized Lotka-Volterra type ordinary differential
equations (ODEs) \cite{Che01}. However, in the absence of interaction
between the nodes, the local dynamics in such a system is trivial.  In
contrast, considering randomly coupled maps as a model for the
dynamical network allows us to consider {\em very general local
  dynamics, including chaos}.  In the specific context of ecological
networks, this is a reasonable assumption for the population dynamics
of individual species. In addition, the use of coupled maps allow us
to work with much larger networks, compared to models incorporating
realistic consumer-resource configurations used to analyze simple
communities with very few species, whose results are difficult to
scale to larger ecosystems \cite{Mcc98}.

Our model has $N$ dynamical elements in a network with random nonlocal
connectivity, for instance representing an ecological network of $N$
interacting species. Each node $i ( = 1 \ldots N)$ is associated with
a continuous state variable $x_n (i)$ which represents the relative
population density of the $i$th species at time $n$. The interaction
between two species is represented by Lotka-Volterra type relation,
with the sign of the coupling coefficient $J_{ij}$ determining either
a predator-prey relation ($J_{ij} > 0, J_{ji} < 0$), competition
($J_{ij}, J_{ji} < 0$) or mutualism ($J_{ij}, J_{ji} > 0$).  The
time-evolution of the system is given by
\begin{equation}
x_{n+1}(i) = f [ x_n (i) \{ 1 +  \Sigma_j J_{ij} x_n(j) \} ],
\end{equation}
where $f$ represents the local on-site dynamics. For the results shown in this
paper we have chosen $f$ to be the exponential map,
\begin{equation}
f (x) = x e^{r(1-x)}, ~{\rm if}~ x > 0;  ~= 0, ~{\rm otherwise}.
\end{equation}
$r$ being the nonlinearity parameter leading from periodic behavior to
chaos \cite{Ric54}. This is a much more realistic model of population
dynamics than the logistic map, and in contrast to the latter, is
defined over the semi-infinite interval $[0, \infty]$ rather than a
finite, bounded interval. Our results also hold for other models of
population dynamics such as the Bellows map, $f (x) = r x / (1 +
x^b)$ \cite{Bel81}.  These maps have the property that they do not go extinct in
the absence of coupling, as we are interested not in intrinsic
instability of the species, but rather in the instability induced by
network interactions.

The connectivity matrix ${\bf J} = \{ J_{ij} \}$ is a sparse matrix
with probability $1-C$ that an element is zero. The diagonal entries
$J_{ii} = 0$ indicate that in the absence of interaction with other
species, the exponential map (2) completely determines the population
dynamics of each species. The non-zero entries in the matrix are
chosen from a normal distribution with mean 0 and variance $\sigma^2$.
Note that we have also used uniform distribution over the interval
$[- \sigma, \sigma]$ without any qualitative changes in the results.
The results reported below are for parallel updating; similar results
hold for random sequential updating.  Also, our results hold for
interaction couplings other than the one used above. For example, the
following type of coupling:
$$
x_{n+1}(i) = f [ x_n (i) ] + \Sigma_j J_{ij} f [ x_n(i) ] f [
x_n(j) ],$$
gives results similar to that reported in this paper.

The linear stability criteria for random networks provides a relation
between the parameters $N$, $C$ and $\sigma$. However, since we are
considering explicit local dynamics, we have an additional parameter,
$r$.  In our work, instead of looking at linear stability, we shall
consider persistence, i.e., the probability that a site has a non-zero
value of $x$, as the measure of stability of the system.  Although
some early work on survival and extinction of species in a coupled
network were done in restricted contexts of exclusively competitive
\cite{Abr98} or cooperative interactions \cite{Sta93}, no systematic
study has been previously made on whether the May criterion is valid
in the presence of local dynamics, incorporating all kinds of interactions
between species.

Initially, all the $N$ species have population values randomly
distributed about $x = 1$.  Immediately after starting the simulation
the number of persistent species (i.e., with $x > 0$) decreases
rapidly, but eventually attains a steady state value which is a
function of the system parameters. 
Note that, if $x \le 0$ for any species, it is removed from the system and
subsequently plays no further role.  After a series of such
extinctions, the effective number of interacting species decreases
and, consequently, the intensity of such extinction-inducing
fluctuations is also reduced.  We have continued the simulations for
up to $10^4$ iterations, when the probability of further extinctions
was found to become extremely small. We then look at the fraction of
species which survive as a function of the model parameters (Fig. 1).
The results qualitatively agree with the May criterion for stability,
in that, increasing complexity (in terms of size, connectivity and
interaction strength of the network) decreases stability,
with a larger proportion of species liable to get extinct. Note that
the May criterion was derived on the basis of local stability, whereas
here we are considering the species persistence, a measure of global
stability.

Fig. 1(a) shows the ratio of persistent species $N_{pers}$ with respect to the
initial number of species $N$. This ratio $N_{pers}/N$ appears to
vary as $1/N$ for large $N$. This indicates that
the number of surviving species is independent of $N$.
Agreement with Wigner-May stability results is also
seen for the $1/C$ variation of surviving fraction with connectivity
(Fig. 1(b)).  Fig. 1(c) shows that the fraction of survivors depend on
the interaction strength parameter $\sigma$ as $1/\sigma^z$ where the
exponent $z$ is an increasing function of the connectivity $C$. This
dependence is expected because, if $C$ is decreased while keeping $N$
fixed, the effective number of other species that a species interacts
with, is decreased. In the limit $C \rightarrow 0$, every species is
independent of all other species, and will persist with probability 1.
Finally, we display the survival fraction against the nonlinearity
parameter $r$ of the local map.  It is clearly evident that one
obtains a smooth monotonic variation of the survival fraction with
respect to $r$ (Fig. 1(d)). This a priori may seem surprising, since
the local map has a significant range of diverse dynamics including
windows of periodic and chaotic behavior and this is not reflected at
all in the figure.

To understand these results, we analyze the probability of survival
of any species in the steady state.  A species $i$ will become extinct
if its population $x_i$ becomes negative at a particular time. By
looking at the equations describing the system, one notes that this is
only possible if $\Sigma_j J_{ij} x_j < - 1$. Therefore, the
probability of survival of a species is essentially equivalent to $P
(\Sigma_j J_{ij} x_j > - 1)$. The distribution of $P (\Sigma_j J_{ij}
x_j)$ has a power law distribution about its peak at zero, and
Gaussian tails.
We now scale this distribution with respect to the different 
network parameters, as scaling in non-equilibrium phenomena is 
the most sensitive and stringent test of universality.

Fig. 2 shows the scaling of $P (\Sigma_j J_{ij} x_j)$ with the
connectivity $C$ which goes as $\sim C^{-1} g_c (C^{-\beta} \Sigma_j
J_{ij} x_j)$ where $g_c$ is the scaling function independent of $C$,
implying $C P (\Sigma_j J_{ij} x_j > -1)\sim$ constant. 
Therefore, the probability of survival varies as $\sim 1/C$, in
exact agreement with the results obtained from linear stability
analysis.
The exponent $\beta
= 0.2 \pm 0.02$ for a wide range of values of $\sigma$ and $r$. 
Similar agreement is
seen for the variation of the probability of survival with $\sigma$
(Fig. 3).  The scaling data show that $P (\Sigma_j J_{ij} x_j) \sim
\sigma^{-2} g_{\sigma} (\sigma^{- \alpha} \Sigma_j J_{ij} x_j)$, where
$g_{\sigma}$ is the scaling function, so that the survival
probability varies as $\sim 1/\sigma^2$. The exponent $\alpha$ varies
in the range 0.1--0.2, decreasing with $r$ and with $C$.

The variation with the map nonlinearity parameter however has no
analog in the previous work on random networks. We observe that the
relevant parameter is the image of the critical point of the map,
rather than $r$ itself.  This point $x_r^{max} = e^{(r-1)}/r$ gives a
measure of the width of the chaotic attractor \cite{lyapnote}. 
Since this increases
the interval over which the probability of $(\Sigma_j J_{ij} x_j)$ is
observed, we have normalized the argument of the scaling function by
dividing it by $x_r^{max}$. Fig. 4 shows the scaling of $P (\Sigma_j
J_{ij} x_j) \sim (x_r^{max})^{- \gamma} g_r [(x_r^{max})^{- 1}
\Sigma_j J_{ij} x_j]$, where $g_r$ is the scaling function.
Therefore, the probability of survival varies as
$(x_r^{max})^{- \gamma}$, with the exponent $\gamma = 3.1 \pm 0.1$ for
a wide range of values of $C$ and $\sigma$. Interestingly, when the
local dynamics is given by the Bellows map, we again obtain $\gamma
\sim 3$.

The above scaling results show that the complexity-stability relations
obtained by May hold true not only qualitatively, but also
quantitatively, when we introduce explicit local dynamics of the
network elements.  The exact nonlinearity of the map, as would be
reflected in, e.g., the Lyapunov exponent, does not enter any of the
results, which suggests that these relations are universal and
independent of details of the local dynamics. In addition, the results
remain valid even if the local nonlinearity parameter $r$ for all the
$N$ maps is not a constant, but varies according to a uniform
random distribution between $r = 2$ and $r = 4$.

The power spectra of quantities such as the total system population,
$\sum_{i=1}^N x_i$ (which can be identified with ``biomass'' in the
ecological context), has a low frequency scaling given by : $S(f) \sim
f^{-\alpha}$ with $1 < \alpha < 2$. In addition, the distribution of
populations $P(x)$ is a clear power law: $P(x) \sim x^{-\phi}$, with
$\phi \sim 1$ for sufficiently high $r$ [Fig. 4 (inset)] \cite{note1}.  

In summary, our work addresses one of the strong criticisms against
the wider applicability of the May-Wigner results, namely their assumption
of an equilibrium. Here we have a range of dynamics at the local level
and certainly no dynamical equilibrium at the global level, as
populations are always fluctuating. Rather we have a non-equilibrium
steady state
where the survival
fraction attains stationarity. The stability of our dynamically more
complex network however still obeys the May criterion, and increasing
complexity (in terms of size, connectivity and interaction strength of
the network) leads to greater instability, resulting in a larger
proportion of species becoming extinct \cite{note0}.  
Scaling results of the
probability distribution of the interaction term in the stationary state 
indicate that the stability of the network 
varies as $\sim \frac{1}{N C \sigma^2}$, 
very much in agreement with the May-Wigner results. We
also find that the stability of the network scales with the
nonlinearity parameter of the local maps in a smooth monotonic
fashion, with the relevant scaling variable being the maximum value
that $x$ can take (which depends monotonically on the nonlinearity).
These observations hold for networks with widely varying local
dynamics as well as for different updating and coupling schemes,
underscoring a remarkable universality and increasing the scope of
relevance of the May-Wigner stability theorem.

{\small We thank Prashant Gade, Sanjay Jain, Purusattam Ray, Somdatta Sinha
and Chris Wilmers for helpful discussions.
This research was supported in part by the National Science Foundation 
under Grant No. PHY99-07949.}

\begin{figure}[t!]
\centerline{\includegraphics[width=0.85\linewidth,clip] {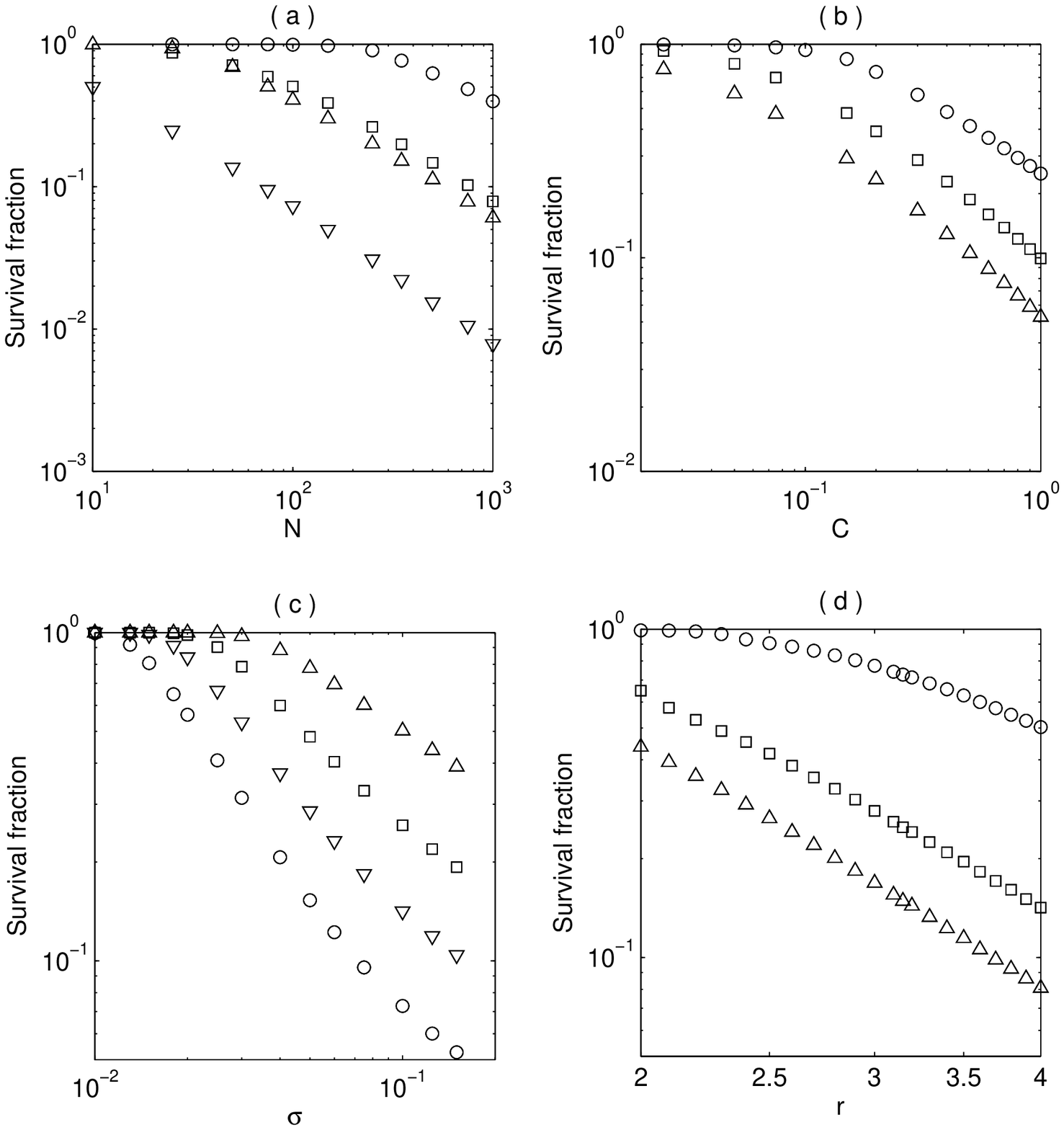}}
\vspace{0.5cm}
\caption{\small The fraction of persistent nodes plotted against the model
parameters: (a) the initial number of nodes, $N$ ($\sigma = 0.1$;
$\circ$: $C = 0.1, r = 2$, $\Box$: $C = 0.1, r = 4$,
$\bigtriangleup$: $C = 1, r = 2$, $\bigtriangledown$: $C = 1, r = 4$;
(b) connectivity, $C$ ($N = 100$, $\sigma = 0.15$; $\circ$: $r = 2$,
$\Box$: $r = 3$, $\bigtriangleup$: $r = 4$);
(c) standard deviation, $\sigma$ ($N = 100$, $r = 4$; $\bigtriangleup$:
$C = 0.1$, $\Box$: $C = 0.25$, $\bigtriangledown$: $C = 0.5$, $\circ$:
$C = 1$); (d) the nonlinearity
parameter, $r$ ($N = 100$, $\sigma = 0.1$; $\circ$: $C = 0.1$, $\Box$:
$C = 0.5$, $\bigtriangleup$: $C = 0.9$). The data is obtained
after $10^4$ iterations and averaged over 5000 realizations.}
\end{figure}

\begin{figure}[t!]
\centerline{\includegraphics[width=0.95\linewidth,clip] {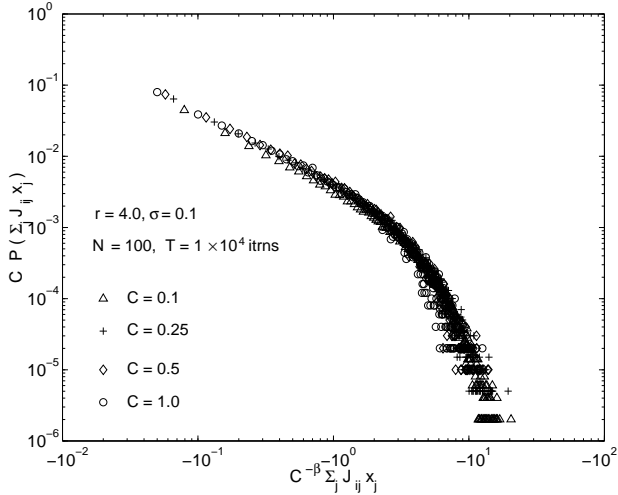}}
\vspace{0.5cm}
\caption{\small The scaling of $\Sigma_j J_{ij} x_j$ with
connectivity $C$ for $N = 100$, $\sigma = 0.1$ and $r = 4$.
The data is obtained
after $10^4$ iterations and averaged over 5000 realizations.}
\end{figure}

\begin{figure}[t!]
\centerline{\includegraphics[width=0.95\linewidth,clip] {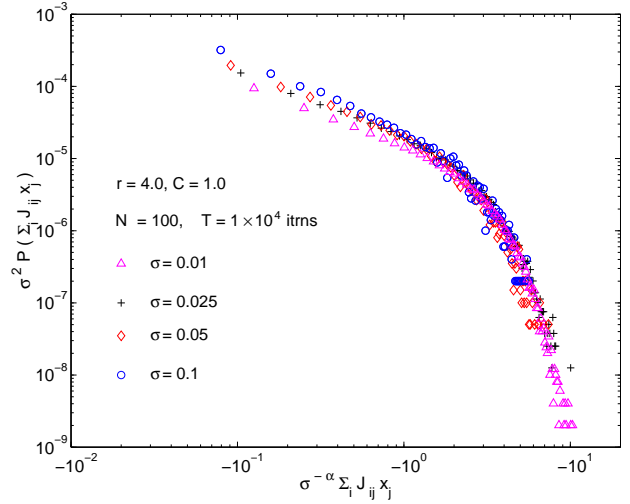}}
\vspace{0.5cm}
\caption{\small The scaling of $\Sigma_j J_{ij} x_j$
with $\sigma$, the standard deviation
of normal distribution from which the connection weights are chosen
($N = 100$, $C = 1$ and $r = 4$). The data is obtained
after $10^4$ iterations and averaged over 5000 realizations.}
\end{figure}

\begin{figure}[t!]
\centerline{\includegraphics[width=0.95\linewidth,clip] {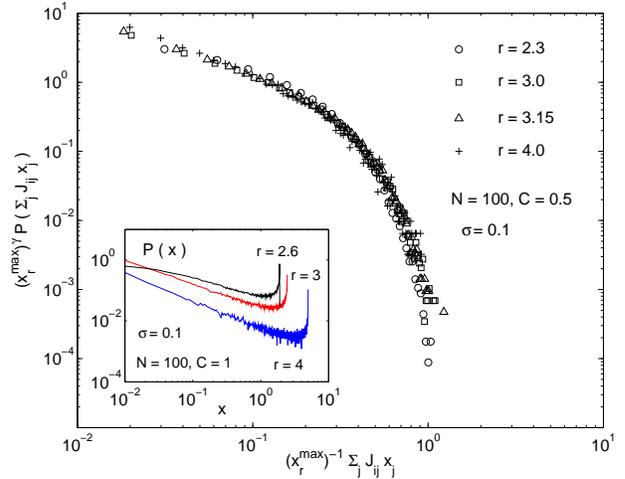}}
\vspace{0.5cm}
\caption{\small The scaling of  $\Sigma_j J_{ij} x_j$
with the width of the attractor $x^{max}_r$ for $N = 100$, $C = 0.5$
and $\sigma = 0.1$.
The inset shows the power-law scaling behavior of the probability
distribution of populations $x$ ($N = 100$, $C = 1$ and $\sigma = 0.1$).
The data is obtained
after $10^4$ iterations and averaged over 5000 realizations.}
\end{figure}


\vspace{-0.5cm}
\begin{thebibliography}{10}
\vspace{-1.5cm}
\bibitem{Dro02}
 B. Drossel and A. J. McKane, nlin.AO/0202034 (2002).

\bibitem{May73}
 R. M. May, {\em Stability and Complexity in Model Ecosystems} (Princeton
 Univ. Press, Princeton, NJ, 1973).

\bibitem{Meh91}
 M. L. Mehta, {\em Random Matrices} (Academic Press, San Diego,
 2nd ed., 1991).

\bibitem{May72}
 R. M. May, Nature {\bf 238}, 413 (1972).

\bibitem{Has82}
 H. M. Hastings, J. Theo. Biol. {\bf 97}, 155 (1982).

\bibitem{Gar70}
 M. R. Gardner and W. R. Ashby, Nature {\bf 228}, 794 (1970).

\bibitem{McC00}
 K. S. McCann, Nature {\bf 405}, 228 (2000).

\bibitem{Pim84}
 S. L. Pimm, Nature {\bf 307}, 321 (1984).

\bibitem{Wil02}
 C. C. Wilmers, S. Sinha and M. Brede, Oikos {\bf 99}, 363 (2002).

\bibitem{Maj99}
 S. N. Majumdar, Curr. Sci. {\bf 77}, 370 (1999).

\bibitem{Rag95}
 S. Raghavachari and J. A. Glazier, Phys. Rev. Lett. {\bf 74}, 3297 (1995).

\bibitem{Wat02}
 D. J. Watts, Proc. Natl. Acad. Sci. USA {\bf 99}, 5766 (2002).

\bibitem{Che01}
 X. Chen and J. E. Cohen, J. Theo. Biol. {\bf 212}, 223 (2001).

\bibitem{Mcc98}
 K. S. McCann, A. Hastings and G. R. Huxel, Nature {\bf 395}, 794 (1998).

\bibitem{Ric54}
 W. E. Ricker, J. Fish. Res. Board Can. {\bf 11}, 559 (1954);
 M. P. Hassell, J. H. Lawton and R. M. May, J. Anim. Ecol. {\bf 45},
 471 (1976);
 S. Sinha and S. Parthasarathy, Proc. Natl. Acad. Sci. USA {\bf 93},
 1504 (1996).

\bibitem{Bel81}
 T. S. Bellows, J. Anim. Ecol. {\bf 50}, 139 (1981).

\bibitem{Abr98}
 G. Abramson and D. H. Zanette, Phys. Rev. E {\bf 57}, 4572 (1998).

\bibitem{Sta93}
 P. F. Stadler and R. Happel, Math. Biosci. {\bf 113}, 25 (1993)

\bibitem{lyapnote}
 We computed the lyapunov spectra of the system for various
 parameter values to verify that the global dynamics showed 
 strong spatiotemporal chaos with no apparent synchronization
 between nodes.
 

\bibitem{note1}
An additional feature of
the distribution is that it is bounded precisely at $x_{max}$, and
there is a buildup of probability at $x \rightarrow x_{max}$.
Interestingly, this kind of distribution, namely power law scaling at
the lower end and an enhanced probability at the outer bound of the
distribution, is seen in critical states arising in networks of
nonlinear maps under threshold activated coupling. See, e.g.,
S. Sinha and D. Biswas, Phys. Rev. Lett. {\bf 71} (1993) 2010.

\bibitem{note0}
 The results hold qualitatively even when the connectivity matrix
 has a small-world topology. See, e.g., S. Sinha, Physica A {\bf 346}
 (2005) 147.

\end{thebibliography}
\end{document}